\documentclass[12pt]{article}
\usepackage{graphicx}

\newcommand{\nn}{\nonumber}

\newcommand{\be}{\begin{equation}}
\newcommand{\ee}{\end{equation}}
\newcommand{\ba}{\begin{eqnarray}}
\newcommand{\ea}{\end{eqnarray}}
\newcommand{\bdm}{\begin{displaymath}}
\newcommand{\edm}{\end{displaymath}}

\def\bea{\begin{eqnarray}}
\def\eea{\end{eqnarray}}
\def\de{\partial}

\def\sla{\raise.15ex\hbox{$/$}\kern-.57em}
\def\ie{{\it i.e.}~}

\def\th{\theta}

\def\f{\phi}

\def\w{\omega}

\def\cG{{\cal G}}

\def\cL{{\cal L}}

\def\cN{{\cal N}}

\def\cY{{\cal Y}}

\begin{document}

\begin{titlepage}

\begin{center}
{\Large\bf {Motivations for AdS/QCD from 10D supergravity solutions.}}
\end{center}

\begin{center}
\bf{Wayne de Paula}\\
{\sl Dep. de F\'\i sica, Instituto Tecnol\'ogico de Aeron\'autica, CTA, 12228-900 \\ S\~ao Jos\'e dos Campos, Brazil}
\end{center}


\begin{abstract}
We discuss some attempts for the construction of gravity duals of QCD-like theories. It is analysed some properties of solutions of 10D Type IIB supergravity theory that attempt to be dual to $\cN$=1 gauge theories, in particular the solutions that belong to Papadoulos-Tseytlin ansatz. We argue that one could obtain 5D effective theories from 10d solutions and it motivates the use of phenomenological AdS/QCD models.
\end{abstract}

\end{titlepage}

\section{Introduction}
Due to the non-linear structure of QCD, most analytical techniques rely on perturbation theory. For very high energies, asymptotic freedom ensures a small coupling constant, that allows one to use perturbative methods to obtain physical amplitudes. In opposition, for low energies, QCD is strongly coupled and therefore the perturbative expansion is meaningless. This makes it very difficult to find analytical tools to analyze the low-energy sector of QCD. This is the main reason why important properties of strong interactions associated with the infrared (IR) physics such as confinement, mass gap and linear Regge trajectories remains unexplained from the QCD Lagrangian using analytical methods.

Nowadays, different techniques have been developed to study non-perturbative aspects of QCD. Examples of such methods are QCD sum rules\cite{QCD_sum_rules_shifman,QCD_sum_rules_review,marina}, Dyson-Schwinger equations\cite{dyson-schwinger_robert,dyson-schwinger_alkofer} and lattice QCD\cite{Lattice,Orlando}, that requires massive numerical computations.

A remarkable contribution was made by Juan Maldacena in $1998$. He proposed an exact mappping between a supersymmetric gauge theory, $\cN=4$ super Yang-Mills theory, in 4D flat space and Type IIB string field theory in 10D space-time $AdS_{5}\times S_{5}$.

The most interesting fact of this duality is that the strong-coupling regime of large-$N$ gauge theories can be approximated by weakly coupled classical gravities, and vice-versa. Thus, one could use weak-coupling perturbative methods in one theory to investigate the strongly coupled dual theory. As QCD is a gauge theory, we can say that this duality gives some hope for a better understanding of the non-perturbative regime of strong interactions.

Here we are interested in discussing some attempts for the construction of gravity duals of QCD-like theories. In particular, we study 10D solutions of Type IIB supergravity fields that attempt to be dual to $\cN$=1 gauge theories\cite{waynebianchi}. Different from $\cN$=4 super Yang-Mills, $\cN=1$ gauge theory encodes some crucial properties of the strong interaction, such as confinement. Important type IIB solutions as Klebanov-Strassler\cite{Klebanov:2000hb}, Klebanov-Tseytlin\cite{Klebanov:2000nc} and Maldacena-Nunez\cite{Maldacena:2000yy} are candidates of $\cN$=1 holographic dual. They belong to a more general class of solutions proposed by Papadopoulos and Tseytlin\cite{Papadopoulos:2000gj} (PT ansatz). In order to analyze the possible duality between the solutions \cite{Klebanov:2000hb,Klebanov:2000nc,Maldacena:2000yy} and $\cN$=1 SYM, we calculate the isometries of the PT background. We propose an ansatz for the vector fluctuations of the metric and p-forms. Using this ansatz we calculated the spectrum of vector fluctuations for the Maldacena-Nunez solution\cite{Maldacena:2000yy} and found no mass gap \cite{waynebianchi}.

We discuss that 10D solutions have a 5D effective representation and therefore we can adopt an alternative strategy, focusing on 5D gravity model to describe hadronic physics. In this 5D approach we use QCD properties to construct holographic models in a bottom-up way as the Hard Wall\cite{hw}, the Soft Wall\cite{Karch} and the Dynamical AdS/QCD model\cite{dePaulaPRD09}.

\section{10D Supergravity Solutions}

The low-energy limit of string theory is supergravity. Also, if the curvature radius R of the space-time metric is much greater than the string length l$_{s}$ (low-curvature regions), we can approximate the string action by a classical one. As the AdS/CFT duality is proposed for Type IIB superstring theory, it is an important step to analyze solutions of Type IIB supergravity.

A crucial fact for the duality is that $\cN$ = 4 SYM is a conformal field theory (CFT). On the gravity side, it is related to the conformal symmetry presented in the AdS$_{5}$ metric\cite{Isometries}. In more details, in \cite{Kawai_Suyama} is shown that the invariance under a scale transformation of the AdS$_{5}$ leads to a correspondence between Wilson Loops in $\cN$ = 4 SYM and minimal surfaces in AdS$_{5}$, which is one of the claims of the AdS/CFT correspondence\cite{Rey,MaldaPRL}.

A first step in this research program is to find similar correspondences for super Yang-Mills theories with a small supersymmetry group that can incorporate some QCD features as confinement and chiral symmetry breaking, which is not possible in the case of $\cN=4$ SYM. The $\cN=1$ SYM presents some of those properties and therefore it is a supersymmetric theory closer to phenomenology than $\cN=4$ SYM. We will discuss some classical solutions of type IIB supergravity that are proposed duals of $\cN=1$ SYM.

\subsection{Field equations}

In the Einstein frame, the 10d Type IIB supergravity equations read
\bea
 &&R_{MN} = {1\over 2} \de_M \phi \de_N \phi + {1\over 2} e^{2\phi} \de_M \chi \de_N \chi
 +{1\over 96}  G_{MPQKL}G_N{}^{PQKL} + {1\over 4} e^\phi \hat{F}_{MPQ}\hat{F}_N{}^{PQ} \nn\\
 && + {1\over 4} e^{-\phi} H_{MPQ}H_N{}^{PQ} - {1\over 48} g_{MN}
[e^\phi \hat{F}_{LPQ}\hat{F}^{LPQ} + e^{-\phi} H_{LPQ}H^{LPQ}], \label{Ricci}\\
 &&\nabla^2\phi  = e^{2\phi} \de_M \chi \de^M \chi
 + {1\over 12}
e^\phi  \hat{F}_{LMN}\hat{F}^{LMN} - {1\over 12} e^{-\phi} H_{LMN}H^{LMN}, \label{dilaton_eq_motion}\\
&&\nabla^M(e^{2\phi}\de_M\chi)  = - {1\over 6} e^{\phi} H_{LMN}\hat{F}^{LMN},\\
&&\nabla^M(e^{\phi}\hat{F}_{MNP})  = {1\over 6} {G}_{NPQRS}H^{QRS},\label{F3_eq}\\
&&\nabla^M(e^{-\phi}H_{MNP} - e^{\phi}\chi \hat{F}_{MNP})= - {1\over 6} {G}_{NPQRS}{\hat{F}}^{QRS},\label{H3_eq}
\eea
where $\nabla^{M}$ is the covariant derivative. $R_{MN}$ is the Ricci Tensor in ten dimensions, $\phi$ (dilaton) and $\chi$ (axion) are scalar fields. Also, $\hat{F_{3}}$, $F_{3}$, $H_{3}$ and $G_{5}$ are anti-symmetric tensor fields that we represent by differential forms. The auxiliary field is given by:
\be \hat{F}_3 = F_3- \chi H_3, \qquad F_1 = d\chi, \qquad F_3 = dC_2, \qquad H_3 = dB_2. \ee

\section{Papadopoulos-Tseytlin Ansatz}

A consistent truncation of 10-D Type IIB supergravity was found by Papadopoulos and Tseytlin\cite{Papadopoulos:2000gj}. It is based on the ansatz presented in the following:

\begin{itemize}
\item{\textbf{Metric}
\bea ds_{10}^2 &=& e^{2p(z)-{x}(z)}\left(e^{2A(z)}
dx\cdot dx + N_{5} dz^2\right) + N_{5}[e^{{x}(z)+g(z)}\left(e_1^2 + e_2^2\right) \nonumber \\
&&+ {1\over4}e^{{x}(z)-g(z)}\left(\tilde\omega_{1}^2 +
\tilde\omega_{2}^2\right)+
{1\over4}e^{-6p(z)-{x}(z)}\tilde\omega_{3}^2], \nonumber\\
x^{A} &=& \{x^{1}, x^{2}, x^{3}, x^{4}, z,\theta, \tilde\theta, \varphi, \tilde\varphi, \psi\}, \label{metric_PT}
\eea where $z$ denotes the holographic radial coordinate. Note that the metric is written in a diagonal form. This is possible because we used a particular notation, based on the following one-forms.
\bea &&e_{1}
= d\th \quad, \quad
 e_{2} = -\sin \th d\varphi, \nn \\
&&\tilde\omega_{1} = \omega_{1} - a(z) e_{1} \quad, \quad
\tilde\omega_{2} = \omega_{2} - a(z) e_{2}\quad, \quad
\tilde\omega_{3} = \omega_{3} - \cot \theta e_{2}, \nn \\
&& \omega_{1} = \sin \psi \sin \tilde\th d\tilde\varphi + \cos
\psi d\tilde\th \quad , \quad \omega_{2} = - \sin \psi d\tilde\th
+ \cos \psi \sin \tilde\th d\tilde\varphi, \nn \\
&&\omega_{3} = d\psi + \cos \tilde\th d\tilde\varphi,\label{def_forms}
 \eea
where $a(z)$ is a function of the holographic coordinate.
}
\end{itemize}

\begin{itemize}
\item{\textbf{Dilaton, axion, C$_{2}$, B$_2$ and G$_5$}
\bea \phi &=& \phi(z) \quad ,\quad \chi=0, \nonumber\\
\quad C_2 &=& {N_5\over 4} [\psi (e_1\wedge e_2
+ \omega_{1}\wedge \omega_{2}) + b(z) (\omega_{1}\wedge e_1 +
\omega_{2}\wedge e_2)+\cos \theta \cos \tilde\theta d\varphi\wedge d\tilde\varphi],\nonumber\\
B_2 &=& h_{1}(z) (e_1\wedge e_2 + \omega_{1}\wedge \omega_{2}) +
h_{2}(z) (\omega_{1}\wedge e_2 - \omega_{2}\wedge e_1) \nn \\
&&+ h_{3}(z) (-\omega_{1}\wedge \omega_{2} + e_1\wedge e_2),\nn\\
G_5 &=& \cG_5 + * \cG_5 \quad, \quad \cG_5 = K(z) e_1\wedge e_2\wedge \omega_{1}\wedge\omega_{2}\wedge \omega_{3}.
\eea

}
\end{itemize}

Using the PT ansatz we can find type IIB classical solutions. For this we have to solve the coupled system equations (\ref{Ricci}) - (\ref{H3_eq}) using the ansatz proposed by Papadopoulos and Tseytlin for the metric, C$_{2}$, B$_{2}$ and G$_{5}$. The outcome is a system of differential equations in terms of radial variables. In particular, the models proposed by Klebanov-Strassler \cite{Klebanov:2000hb}, Klebanov-Tseytlin \cite{Klebanov:2000nc} and Maldacena-Nunez \cite{Maldacena:2000yy} are solutions that can be cast in a PT format.

\subsection{Isometries}
The Lie derivative of the metric with respect to a Killing vector ($\xi$) is zero, \ie, $\cL_\xi(g_{MN})=0$. Then, we have
\be
\cL_\xi g_{MN} = \xi^L\de_L g_{MN} + g_{ML}\de_{N}  \xi^L + g_{LN}\de_{M}  \xi^L = 0.\label{Lie_der_metric}
\ee
Solving Eq.(\ref{Lie_der_metric}), we found that the Killing vectors of the PT ansatz are
\bea \xi_{+} \equiv \xi_{1} &=& e^{i\varphi} \left(0,0,0,0,0,-1, 0, -i \cot\theta, 0, {i\csc\theta} \right), \nonumber \\
\xi_{-} \equiv \xi_{2} &=& e^{-i\varphi} \left(0,0,0,0,0,1, 0, -i \cot\theta, 0, {i\csc\theta} \right), \nonumber \\
\xi_{3} \equiv \xi_{3} &=& \left(0,0,0,0,0,0,0,1,0,0\right), \nonumber \\
\tilde\xi_{+} \equiv \xi_{4} &=& e^{i\tilde\varphi} \left(0,0,0,0,0,0,-1, 0, -i \cot\tilde\theta, {i\csc\tilde\theta} \right), \nonumber \\
\tilde\xi_{-} \equiv \xi_{5} &=& e^{-i\tilde\varphi} \left(0,0,0,0,0,0,1, 0, -i \cot\tilde\theta, {i\csc\tilde\theta} \right), \nonumber \\
\tilde\xi_{3} \equiv \xi_{6} &=& \left(0,0,0,0,0,0,0,0,1,0\right).\label{Killing_Vectors}
\eea
See that $\xi_a$ have only components in the internal
coordinates, \ie $\xi^M = \delta^M_i \xi^i$ with $M=1,...,10$ and
$i=6,...,10$, and the contra-variant components displayed above
only depend on the internal `angular' variables, \ie $\de_\mu
\xi^M = 0$ with $\mu=1,...,5$. An interesting result is that the Killing vectors of the PT ansatz generates an $SU(2)\times \widetilde{SU(2)}$ algebra. Below we substantiate this claim in detail.

\subsection{$SU(2)\times \widetilde{SU(2)}$ algebra}

Take the Killing vectors $\xi_{a}$ (\ref{Killing_Vectors}) and let us define the following linear combination:
\bea
\varsigma_{1} &=& \frac{1}{2}\left(\xi_{+} + \xi_{-}\right),\quad \varsigma_{2} = \frac{1}{2i}\left(\xi_{+} - \xi_{-}\right),\quad \varsigma_{3} = \xi_{3}, \nn
\eea
that in detail reads:
\bea
\varsigma_{1} &=& \left(0,0,0,0,0,\sin\varphi, 0, \cot\theta \cos\varphi, 0, -\frac{\cos\varphi}{\sin\th} \right), \nonumber \\
\varsigma_{2} &=& -\left(0,0,0,0,0,\cos\varphi, 0, -\cot\theta \sin\varphi, 0, \frac{\sin\phi}{\varphi}\right), \nonumber \\
\varsigma_{3} &=& \left(0,0,0,0,0,0,0,1,0,0\right).
\eea
The generators of the algebra are given in terms of the $3$ Killing vectors, $\varsigma_{a}$, $a = \{1,2,3\}$ as:
\be
L_{1} = -i \varsigma^{\mu}_{1}\partial_{\mu},\quad L_{2} = -i \varsigma^{\mu}_{2}\partial_{\mu},\quad
L_{3} = -i \varsigma^{\mu}_{3}\partial_{\mu},
\ee
that in detail are:
\bea
L_{1} &=& -i\left(\sin\varphi\frac{\partial}{\partial \th} +\frac{\cos\th ~ \cos\varphi}{\sin\th} \frac{\partial}{\partial \varphi} - \frac{\cos\varphi}{\sin\th}\frac{\partial}{\partial \psi}\right),\nn\\
L_{2} &=& i\left(\cos\varphi\frac{\partial}{\partial \th} - \frac{\cos\th ~ \sin\varphi}{\sin\th} \frac{\partial}{\partial \varphi} + \frac{\sin\varphi}{\sin\th}\frac{\partial}{\partial \psi}\right),\nn\\
L_{3} &=& -i \frac{\partial}{\partial \varphi}.
\eea
We can show that the generators above satisfy an SU(2) algebra:
\be
[L_{i},L_{j}] = -i \epsilon_{ijk} L_{k}.\label{SU_2}
\ee
To illustrate that we calculate explicitly $[L_{1},L_{3}]$ and show that it is equal to $-i\epsilon_{132}~L_{2}$. The successive application of the angular operators give:
\bea
L_{1}~L_{3} &=& -\left(\sin\varphi \frac{\partial^{2}}{\partial \th \partial \varphi} - \frac{\cos\varphi}{\sin\th} \frac{\partial^{2}}{\partial \psi \partial \varphi} + \frac{\cos\th \cos\varphi}{\sin\th} \frac{\partial^{2}}{\partial \varphi^{2}} \right),\nn\\
L_{3}~L_{1} &=& -\left(\cos\varphi \frac{\partial}{\partial \th} + \sin\varphi\frac{\partial^{2}}{\partial \th \partial \varphi} + \frac{\sin\varphi}{\sin\th} \frac{\partial}{\partial \psi} - \frac{\cos\varphi}{\sin\th}\frac{\partial^{2}}{\partial\psi\partial\varphi} \right.\nn\\
&&\left.-\frac{\cos\th \sin\varphi}{\sin\th}\frac{\partial}{\partial \varphi} + \frac{\cos\th \cos\varphi}{\sin\th}\frac{\partial^{2}}{\partial \varphi^{2}}
\right),\nn\\
L_{1}~L_{3} - L_{3}~L_{1} &=& \left(-\cos\varphi\frac{\partial}{\partial \th} + \frac{\cos\th ~ \sin\varphi}{\sin\th}\frac{\partial}{\partial \varphi} - \frac{\sin\varphi}{\sin\th}\frac{\partial}{\partial\psi}\right),
\eea
which is equal to $-i\epsilon_{132}~L_{2}$.

Similarly, we can perform the same procedure in relation to the Killing vectors $\tilde\xi_{+}$, $\tilde\xi_{-}$ and $\tilde\xi_{3}$, defining the linear combinations
\bea
\tilde\varsigma_{1} = \frac{1}{2}\left(\tilde\xi_{+} + \tilde\xi_{-}\right),\quad \tilde\varsigma_{2} = \frac{1}{2i}\left(\tilde\xi_{+} - \tilde\xi_{-}\right), \quad \tilde\varsigma_{3} = \tilde\xi_{3}.
\eea
The generators of the algebra are given in terms of the $3$ Killing vectors, $\tilde\varsigma_{a}$, $a = \{1,2,3\}$ as:
\be
\tilde{L}_{1} = -i \tilde\varsigma^{\mu}_{1}\partial_{\mu},\quad \tilde{L}_{2} = -i \tilde\varsigma^{\mu}_{2}\partial_{\mu},
\quad \tilde{L}_{3} = -i \tilde\varsigma^{\mu}_{3}\partial_{\mu}.
\ee
The generators above satisfy an SU(2) algebra:
\be
[\tilde{L}_{i},\tilde{L}_{j}] = -i \epsilon_{ijk} \tilde{L}_{k}.\label{SU_2}
\ee

We can show that not only the metric but also all the $p$-forms of the PT ansatz are invariant under the $SU(2)\times \widetilde{SU(2)}$ isometry generated by the six Killing vectors $\xi^{\mu}_a$, with $(a = 1,...,6)$. Therefore the $SU(2)\times \widetilde{SU(2)}$ isometry is an Exact Symmetry of the theory. In our interpretation, the presence of such symmetry is a remnant of the breaking of $\cN=4$ to $\cN=1$\cite{waynebianchi}.

\section{Maldacena-Nunez Solution}
The Maldacena-Nunez (MN) solution \cite{Maldacena:2000yy} for the coupled type IIB supergravity equations is obtained for the case where G$_5$=0,  H$_3$ =0, $\chi = 0$ and $b=a$. It is given by the metric
\be
ds^{2} = e^{\phi\over 2} \left[dx^2 + N_{5}\left\{dz^2 + e^{2g}\left(e_{1}{}^{2}+e_{2}{}^{2}\right) + {1\over 4}\left(\tilde\w_{1}{}^{2} +
\tilde\w_{2}{}^{2} + \tilde\w_{3}{}^2 \right)\right\} \right],\label{MN_invariant}
\ee
with the radial functions given by
\begin{eqnarray}
&&e^{-2\f} = {2 e^{g} \over \sinh 2z} \quad , \quad
e^{2g} = z ~ \coth 2z - {1 \over 4} (1 + a^2) \nonumber \\
&& a = {2 z \over \sinh 2z} \quad , \quad
f =  4 e^{2g}+a^2. \nonumber \\
\end{eqnarray}

For later use, notice that
\bea && g_{\mu\nu} = e^{\phi \over 2} \delta_{\mu\nu} \quad , \quad g_{ij} = e^{\phi \over 2} \hat{g_{ij}} \quad ,
\quad ||g|| = {N_{5}^{6}\over 64} e^{5\f}e^{4g}\sin^{2}\th ~ \sin^{2}\tilde\th. \nn
\eea

The next step is to define our ansatz for the vector fluctuations of the metric and p-forms in the MN background. We defined a one-form $\mu_1^{\xi}$ such that
\be
i_{\xi}F_{3} = d\mu^{\xi}.
\ee
The outcome for the $\widetilde{SU(2)}$ sector ($\xi_{(4)}$, $\xi_{(5)}$ and $\xi_{(6)}$) of the MN background is given by
\be
\mu_M^\xi = e^{-\phi/2} \xi_M.
\ee

We can define the fluctuations of the metric and p-forms as
\be
\delta g^{MN} = A^M \xi^N + A^N \xi^M \quad ,
\quad \delta C_{MN} = A_M \mu_N^{\xi}- A_N \mu_M^{\xi}\label{ansatz_MN}
\ee
Setting $\delta\phi = 0$, $\delta g_{\mu\nu} = 0, \delta g_{ij} =0$ one
has  $g^{MN} \delta g_{MN} = 0$, \ie $\delta \sqrt{||g||} = 0$. Therefore
\be\delta g_{\mu i} = -A_\mu \xi_i =\delta g_{i\mu}, \quad \delta C_{\mu i} = A_\mu \mu^\xi_i = - \delta C_{i\mu},
\ee and
\bea
 \delta F_{ijk} &=& 0, \quad \delta F_{\mu\nu\rho} = 0, \quad
 \delta F_{\mu ij} = -A_{\mu} \left( \partial _{i} \mu^\xi_{j} - \partial _{j} \mu^\xi_{i}\right), \nonumber \\
 \delta F_{\mu \nu i} &=& \left(\partial_{\mu} A_{\nu} -
\partial_{\nu} A_{\mu} \right)\mu^\xi_{i} -
 \left(A_{\mu} \partial _{\nu} \mu^\xi_{i} - A_{\nu} \partial _{\mu} \mu^\xi_{i}\right).
\eea

\subsection{Spectrum of vector fluctuations}

After the definition of the fluctuations for $\delta g^{MN}$ and $\delta C_{MN}$, we have to apply our ansatz (\ref{ansatz_MN}) for the Type IIB equations. Focusing on the $\widetilde{SU(2)}$ sector of the MN background, we obtain a single dynamical equation:
\be
{1\over \sqrt{g}} \de_\mu [\sqrt{g}e^{\phi/2} f^{\mu\nu}] = 0,\label{massless_vector_MN}
\ee
where $f_{\mu\nu}=\partial_{\mu}A_{\nu} - \partial_{\nu}A_{\mu}$. Note that $f_{\mu\nu}$ is defined in terms of the vector fluctuations $A_{\mu}$ and therefore the solution of the equation of motion (\ref{massless_vector_MN}) gives the vector fluctuations for the $\widetilde{SU(2)}$ sector. There are two cases to consider $\nu = \hat{\nu}$ ($\hat{\nu}$ runs from 1 to 4) and $\nu = 5$. For $\nu = 5$ one simply gets:
\be \de^{\hat\mu} f_{\hat\mu 5} = 0
\quad \rightarrow \quad  \de^{\hat\mu} \de_{\hat\mu} A_5 - \de_5
\de^{\hat\mu} A_{\hat\mu} = 0,
\ee that allows one to express $A_5$ in terms of the longitudinal component of $A_{\hat\mu}$.

For $\nu = \hat\nu$ one gets \be  \de_{\hat\mu}f^{\hat\mu\hat\nu}
+ {1\over \sqrt{g (z)}} \de_5 [\sqrt{g (z)} e^{\phi/2} g^{55}
g^{\hat\nu\hat\lambda} (\de_5 A_{\hat\lambda} - \de_{\hat\lambda}
A_5)] = 0 \ee

Setting $A_{\hat\mu} = a_{\hat\mu}(z) e^{i p\cdot\hat{x}}$ and
$A_{5} = b(z) e^{i p\cdot\hat{x}}$ one can solve for
$a_{\hat\mu}(z)$ and $b(z)$. Decomposing $a_{\hat\mu}(z)$ into
longitudinal ($a_L(z)$) and transverse ($a_{\hat\mu}^T(z)$) components according to
\be
a_{\hat\mu}(z) = a_{\hat\mu}^T(z) + i p_{\hat\mu} a_L(z)
\ee one finds
\be b(z) = a_L'(z),
\ee
which can be set to zero by a gauge transformation. The surviving transverse components then satisfy
an equation
\be {1\over \sqrt{g (z)}} \de_5 (\sqrt{g (z)}
e^{-\phi/2} \de_5 a^T_{\hat\mu}) - e^{-\phi/2}p^2 a^T_{\hat\mu} =0.
\ee
Changing notation $a^T_{\hat\mu} = T$, we have
\be
-p^{2}T + 2 \left(\phi' + g'\right) T' + T''=0
\ee
After setting
\be
T=e^{-\phi-g}\cY
\ee
the equation takes on a Sturm-Liouville form
\be
-\cY'' + p^{2}\cY + \left(\phi''+g''+ \left(\phi'+g'\right)^{2}\right)\cY = 0\label{SL_MN}.
\ee
The first interesting point is that despite the fact that we are solving a 10-dimensional problem, the dynamical fluctuation equation is one-dimensional, in which the only dependence is on the radial coordinate (z). This result motivates one to write an effective Lagrangian in a space with fewer dimensions that gives the same one-dimensional dynamical equation. This relates to the 5-dimensional holographic models that we will discuss in the next section.

The second observation is that there are no discrete states associated to the bulk vector fluctuations for the MN background (\ref{SL_MN}). The reason is that the effective potential given by
\be
V_{eff} = \left(\phi''+g''+ \left(\phi'+g'\right)^{2}\right)
\ee
tends to a constant in the IR limit $(z\rightarrow \infty)$. Despite the area-law behavior of the Wilson loop in the MN background \cite{Maldacena:2000yy}, the continuous spectrum presented by the vector fluctuations disagrees with the holographic interpretation of the MN background as a dual to a confining theory such as $\cN=1$ SYM.

\section{5d Holographic Models}
Starting from a 10D holographic model we can find a 5D effective action. For a review, see e.g. the paper by Berg, Haack and Mueck \cite{Berg:2006xy}. An interesting result of this work is the construction of an effective 5D action with several scalars starting with a 10D consistent truncation of type IIB supergravity described by the PT ansatz\cite{Papadopoulos:2000gj}. In the 5D effective action, each scalar has a different sigma model metric. We can understand this equivalence as a parametrization of the 10D fields of type IIB supergravity by a 5D metric and a set of scalar fields.

In addition to this analysis, another motivation for using the 5D holographic models is the fact that the dynamical equation for vector fluctuations of the MN solution(\ref{massless_vector_MN}) is not dependent on the internal variables. Therefore, from an effective field theory point of view, it would be possible to obtain the same equations of motion neglecting the internal space.

The pioneer 5d holographic model is the Hard Wall\cite{hw} where confinement is introduced by an IR cut-off in the AdS$_{5}$. It can reproduce a huge amount of hadron phenomenology \cite{Teramond_PRL05,Boschi_04,Boschi_06}. In particular, the conformal invariance of AdS$_{5}$ in the UV region reproduces the counting rules which govern the scaling behavior of hard QCD scattering amplitudes\cite{BrodskyFarrar}, while an infrared cutoff on the fifth dimension implements the mass gap and discrete hadron spectra.

However, in contrast to the observed approximate linear Regge behavior\cite{Anisovich_00,Klempt_02,Bugg_04}, the hard-wall predictions for the squared masses of light-flavor hadrons depend quadratically on the radial excitation. In order to obtain linear trajectories, it was proposed a model \cite{Karch} where the AdS$_{5}$ geometry is kept intact while an additional dilaton background field with quadratic dependence on the extra dimension is exclusively responsible for conformal symmetry breaking. However this model does not exhibit the area-law behavior. In addition, the soft-wall background is not solution of the 5D Einstein equations. In the soft wall model the dilaton field is an inert scalar, i.e. it has no dynamics.

We then adopted a phenomenological point of view to access the mesonic spectrum. We found a solution of the five-dimensional Einstein-dilaton equations which provides an approximate dual background for holographic QCD\cite{dePaulaPRD09}. The motivation of this model was to overcome two difficulties presented by the Soft Wall model\cite{Karch}. Using this Dynamical AdS/QCD model we obtained the Regge spectrum of high-spin mesons. Different from the Soft Wall model, the proposed dilaton-gravity background proposed has confinement through Wilson loop analysis\cite{dePaulaNPBPS} and is a classical solution of the 5D Einstein equations.

We applied the ``Dynamical AdS/QCD model" to study scalar and pseudoscalar mesons. We obtained the Regge-like spectrum of the $f_0$ family in agreement with the experimental data\cite{dePaula:2009za}. Using this model we calculated the spectrum of light mesons\cite{dePaulaNPBPS}, including scalars\cite{dePaula:2009za}. We also obtain the decay amplitude of scalars into two pions\cite{dePaulaIJMP}.

\section{Summary}
It is important to stress several differences between the dual particle representation in the 5D models and the 10D models. In the 10D perspective, we interpreted the vector fluctuations of the background of a classical solution (gravity side) as the holographic dual of vector particles (field theory side). On the other hand, in the 5D model, we introduced an additional tensor field of spin S propagating in a deformed AdS$_{5}$ as the dual representation of a spin S meson\cite{dePaulaPRD09}. We believe that the necessity of introducing an additional field, instead of a fluctuation of the background fields, is related to the fact that our 5D model is an ``effective" representation of a 10D solution. Therefore the physical complexity of the 10D tensor fields is naively represented by the dilaton potential of our 5D Dynamical model, V($\Phi$). Following this interpretation, we argue that the additional 5D fields propagating in a deformed AdS$_{5}$ play the role of the vector fluctuation of the metric and p-forms in a 10D model.

We are also interested in modifying the MN solution to introduce $\Lambda_{QCD}$ at some level, in an attempt to obtain a discrete mass spectrum for the fluctuations. The absence of a mass gap in the MN solution is connected to the IR behavior of the effective potential of the corresponding dynamical equation. The challenge would be to find a solution of the type IIB field equations (\ref{Ricci}) - (\ref{H3_eq}) that has a confining effective potential (i.e $V_{eff}\sim z^{\lambda}$, $\lambda>1$) for the fluctuations. This could lead to another step in the bottom-up approach to duality by starting from a 10D perspective to obtain hadronic properties.

Finally, we hope that our phenomenological study of a 5D dynamical model, consistent with the available experimental data, may constitute a useful guidance to building a 10-dimensional gravity model that describes the hadron spectrum. A 10-dimensional supergravity theory dual to QCD is still missing. This is a hopeful goal because, from the supergravity side, one would approach a low energy limit of string theory that aims to describe the strong interaction. The bottom-up approach looks promising from the point of view of the phenomenology, but there is still a long journey to a 10-dimensional theory.

\end{document}